\documentclass[10pt,aps,showpacs,superscriptaddress,prl]{revtex4-1}
\usepackage[caption=false]{subfig}
\usepackage{amsmath}
\usepackage[utf8]{inputenc} 
\usepackage[T1]{fontenc} 
\usepackage{graphicx}
\usepackage{xcolor}
\graphicspath{{./Figures/}}

\begin{document}

\title{Complete elimination of information leakage in continuous-variable quantum communication channels}

\author{Christian S. Jacobsen}
\email{Corresponding author: chrsch@fysik.dtu.dk}
\affiliation{Center for Macroscopic Quantum States (bigQ), Department of Physics, Technical University of Denmark, Fysikvej, 2800 Kongens Lyngby, Denmark}
\author{Lars S. Madsen}
\affiliation{Center for Macroscopic Quantum States (bigQ), Department of Physics, Technical University of Denmark, Fysikvej, 2800 Kongens Lyngby, Denmark}

\author{Vladyslav C. Usenko}
\affiliation{Department of Optics, Palacky University, 17. listopadu 12, 771 46 Olomouc, Czech Republic}

\author{Radim Filip}
\affiliation{Department of Optics, Palacky University, 17. listopadu 12, 771 46 Olomouc, Czech Republic}

\author{Ulrik L. Andersen}
\affiliation{Center for Macroscopic Quantum States (bigQ), Department of Physics, Technical University of Denmark, Fysikvej, 2800 Kongens Lyngby, Denmark}

\date{\today}

\begin{abstract}
In all lossy communication channels realized to date, information is inevitably leaked to a potential eavesdropper. Here we present a communication protocol that does not allow for any information leakage to a potential eavesdropper in a purely lossy channel. By encoding information into a restricted Gaussian alphabet of squeezed states we show, both theoretically and experimentally, that the Holevo information between the eavesdropper and the intended recipient can be exactly zero in a purely lossy channel while minimized in a noisy channel. This result is of fundamental interest, but might also have practical implications in extending the distance of secure quantum key distribution.
\end{abstract}


\maketitle

\section*{Introduction}
\begin{figure}[h]
\includegraphics[width=0.4\textwidth]{
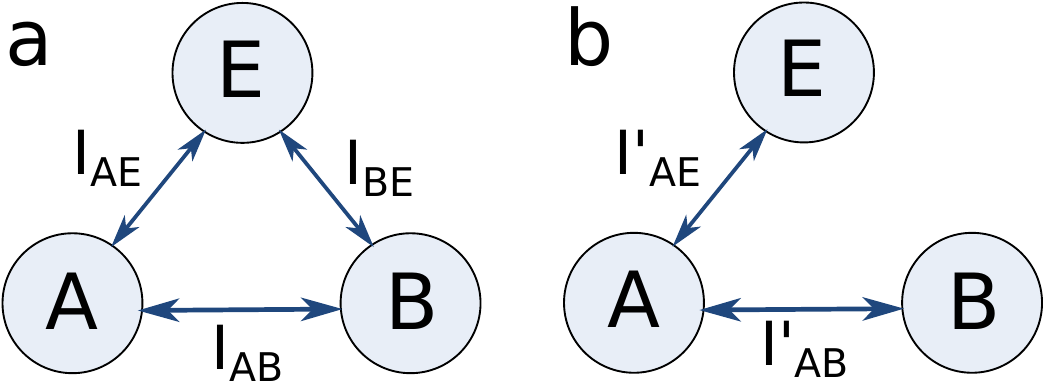}
\caption{A tripartite communication scenario between Alice (A), Eve (E) and Bob (B). a) Each party shares some amount of mutual information, given in terms of Shannon entropies as $I_{XY}=H(X)+H(Y)-H(X,Y)$, with the other two parties. b) In the context of message security, it is the goal of the honest parties, Alice and Bob, to completely eliminate the information that they share with Eve. By removing all correlations between Eve and Bob (that is $I_{BE}^{\prime}=0$), the adversary obtains no information about what Bob has measured, and thus secret communication between A and B can be established.}
\label{fig:Tripartite}
\end{figure}

Security in communication is of utmost importance in modern society. It allows for the delivery of information to the intended recipients while preventing unauthorized eavesdroppers from accessing it. Conceptually, it can be treated as a tripartite communication network in which two entities (e.g. Alice and Bob) intend to communicate while a third party - the eavesdropper (known as Eve) - tries to intercept the message. See Fig.~\ref{fig:Tripartite}, where the mutual information between the three parties is represented schematically. If successful, the interception will generate correlations between all three parties, as in Fig.~\ref{fig:Tripartite}(a), possibly rendering the communication scheme insecure. To regain security, the correlations between the intended recipient and the interceptor must be suppressed. This can be done by means of data post-processing such as privacy amplification - a method commonly used to establish security in quantum key distribution (QKD) schemes~\cite{Gisin2002,Scarani2009}. However, privacy amplification is only successful if the information $I_{AB}$ between the trusted parties Alice and Bob is larger than the information between Bob and Eve prior to the implementation of the procedure~\cite{Devetak2005}.

As an alternative to data postprocessing, the information gained by an eavesdropper can be suppressed by using an entanglement-based protocol followed by entanglement distillation or purification~\cite{Braunstein2005}. Here the two communicating parties seek to share entangled states but due to the interception, the system ends up in a three-party entangled state, subsequently reduced to a purified two-party entangled state between Alice and Bob, thereby eliminating the correlations with the eavesdropper. This strategy is however very challenging as it requires multi-copy non-Gaussian transformations in conventional communication schemes based on Gaussian states encoding and homodyne/heterodyne detection~\cite{Eisert2002, Fiurasek2002, Giedke2002}.

In this Letter we present a completely different approach for minimizing information leakage which is not based on conventional \textit{a posteriori} error correction or privacy amplification and therefore does not rely on any prior information advantage. Instead of suppressing the information of Eve by privacy amplification or distillation at Bob's station, we propose the opposite approach of designing the Gaussian input states and alphabet at Alice's station in such a way that Eve cannot gain {\it any information at any time} in a purely lossy channel, as in Fig.~\ref{fig:Tripartite}(b). We show that by encoding the information into squeezed states of a restricted Gaussian alphabet it is possible to {\it completely and deterministically eliminate the presence of an eavesdropper}, corresponding to the realization of a channel with a Holevo information of zero. The protocol is based on continuous variables (CV) in which quadratures are modulated and measured with homodyne detectors \cite{Braunstein2005, Weedbrook2011, Andersen2010} which is contrasted with discrete variables communication where photon counters are used. We note that no analogue of our proposed scheme for the complete elimination of the Holevo information is known for discrete variables. Unlike covert communication~\cite{Bash2015, Arrazola2016} where the transmission of information is hidden from the eavesdropper, the presence of the signal states are still detectable by Eve in the proposed scheme. In contrast to the private states known from discrete variables~\cite{Horodecki2005, Dobek2011}, which still rely on distillation procedures, our method allows for direct elimination of the information accessible by an eavesdropper using proper state preparation and ideally needs no distillation.

\section*{Results}

We consider the elimination of information leakage in the context of QKD. In CVQKD protocols with reverse reconciliation \cite{Cerf2001, Silberhorn2002, Garcia-Patron2009, Pirandola2008, Grosshans2002, Weedbrook2004, Reid2000, Hillery2000, Ralph2000} the lower bound on the rate of secret key generation in the asymptotic limit of an infinitely long key is given by 
\begin{equation}
R = \beta I_{AB} - \chi_{EB} \ ,
\label{eq:rate}
\end{equation}
where $I_{AB}$ is the mutual information between Alice and Bob as defined through the Shannon entropy \cite{Shannon1948a, Devetak2005}, $\beta \in ]0;1]$ is the post-processing efficiency, and $\chi_{EB}$ is the Holevo information which is an upper bound on the information $I_{EB}$ acquired by Eve \cite{Holevo1973}. A secret key can therefore only be generated when $\beta I_{AB} > \chi_{EB}$. In all previously proposed protocols, the Holevo information has been non-zero (even in principle), which in turn has put stringent conditions onto the processed mutual information between Alice and Bob, $\beta I_{AB}$. This condition has been experimentally fulfilled by applying state-of-the-art postprocessing protocols \cite{Leverrier2008} with high efficiency and low-noise homodyne detectors \cite{Weedbrook2011, Eberle2013, Lance2005, Madsen2012, Grosshans2003, Jouguet2013, Heim2010}. These stringent conditions on Bob's measurements and data processing to enable security can however be largely relaxed by reducing the Holevo information that upper-bounds the information leakage.   

\subsection*{Minimization of information leakage} 
\begin{figure}
\includegraphics[width=0.45\textwidth]{
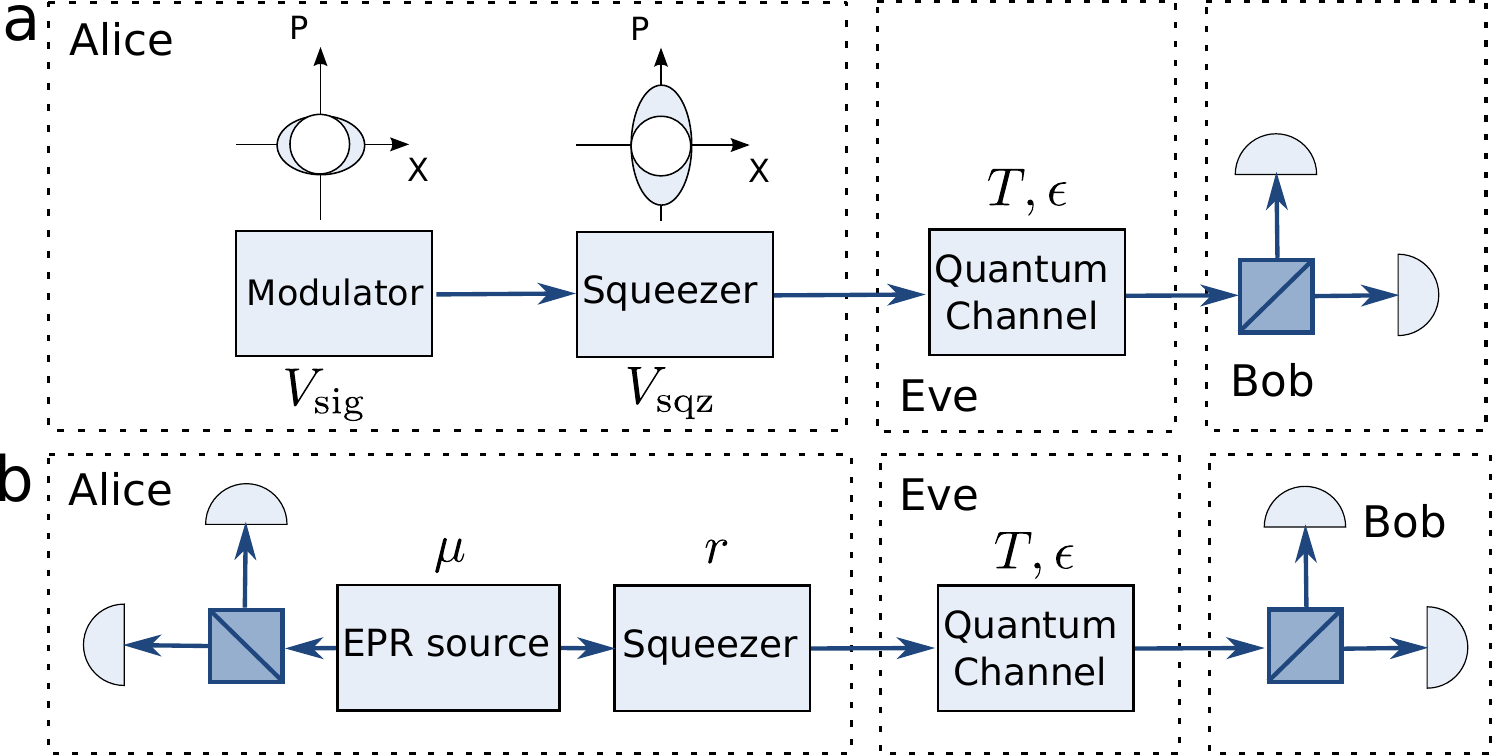}
\caption{Equivalent protocol schemes. (a) Prepare-and-measure scheme for a quantum communication protocol with zero information leakage. An ensemble of amplitude quadrature displaced coherent states is squeezed to have an overall amplitude quadrature noise variance of vacuum before being sent into the quantum channel. (b) Equivalent entanglement-based scheme of the quantum communication protocol with zero information leakage. An EPR state is prepared, with a local mode measured by Alice and the outgoing mode squeezed before being sent into the quantum channel.}
\label{fig:Scheme_Entanglement}
\end{figure}

We consider a prepare-and-measure CVQKD protocol where information is encoded solely into a single quadrature (here the amplitude quadrature $X$ with a variance $V_\text{sig}$) of a Gaussian squeezed state of amplitude quadrature variance $V_\text{sqz}$ (see Fig. \ref{fig:Scheme_Entanglement}(a)), and investigate theoretically under which condition it is possible to completely decouple a potential eavesdropper  from the channel. 

The maximal information, that is the capacity, between Eve and Bob is given by the Holevo quantity: 
\begin{equation}
\chi_{EB} = S(E) - S(E|B) \ ,
\label{eq:Holevo}
\end{equation}
where $S(E)$ is the von Neumann entropy of the state received by Eve and $S(E|B)$ is the von Neuman entropy of the state at Eve conditioned on the measurement at Bob. In the case of a noisy quantum channel, the general collective attack can be accessed by assuming that Eve holds the purification of the state shared between Alice and Bob~\cite{Garcia-Patron2006, Nielsen2010}. Using the triangle inequality, one can derive the self-duality property of the von Neumann entropy, which states that $S(E) = S(AB)$ and $S(E|B) = S(A|B)$~\cite{Araki1970}. From (\ref{eq:Holevo}), it is clear that a Holevo information of zero requires $S(E) = S(E|B)$, and from the purification this translates into $S(AB) = S(A|B)$. This condition is evaluated in the following by individually deducing $S(AB)$ and $S(A|B)$. 

For a Gaussian protocol, where Gaussian attacks are optimal in the asymptotic limit~\cite{Garcia-Patron2006, Navascues2006}, the von Neumann entropies may be easily calculated from the symplectic spectrum of the covariance matrices of the corresponding states~\cite{Weedbrook2011}. To enable an explicit protocol description, we switch to the equivalent EPR based protocol~\cite{Grosshans2003b} where an asymmetric two-mode squeezed state is shared between Alice and Bob as shown in Fig. \ref{fig:Scheme_Entanglement}(b)~\cite{Usenko2015, Gehring2016,Usenko2015,Derkach2016}. 

The variance of a symmetric two mode squeezed vacuum state is denoted $\mu$ while the single mode squeezing transformation is represented by the squeezing parameter $r$ such that amplitude and phase quadrature variances of the modes sent to Bob are $\mu e^{-2r}$ and $\mu e^{2r}$, respectively. The shared state is represented by a covariance matrix, which we may generally write as,

\begin{equation}
\gamma_{AB} = 
\begin{bmatrix}
\gamma_A & \sigma_{AB} \\
\sigma_{AB} & \gamma_B
\end{bmatrix} \ ,
\label{eq:matrix}
\end{equation}
where $\gamma_A = \text{diag}[\mu,\mu]$ is the covariance matrix of the EPR mode kept by Alice, $\gamma_B = \text{diag}[T(e^{-2r} \mu + \epsilon) + 1 - T, T(\mu e^{2r} + V_{\epsilon} + \epsilon) + 1 - T]$ is the covariance matrix of the mode received by Bob, and $\sigma_{AB} = \text{diag}\left[ \sqrt{T e^{-2r}(\mu^2 - 1)},-\sqrt{T (\mu^2 - 1)e^{2r} } \right]$ is the sub-block of the global covariance matrix describing the correlation between modes. Here $T$ is the transmittance, $V_\epsilon$ is the variance of the excess noise of the anti-squeezed quadrature while $\epsilon$ represents the quadrature symmetric excess noise contribution of the channel. $\gamma_{AB}$ is constructed such that the prepare-and-measure scheme, in Fig. \ref{fig:Scheme_Entanglement}(a), and the EPR scheme, in Fig. \ref{fig:Scheme_Entanglement}(b), are equivalent if $\mu=\sqrt{1+V_\text{sig}/V_\text{sqz}}$ and $r=-1/2 \ln[\sqrt{V_\text{sqz}(V_\text{sqz}+V_\text{sig})}]$~\cite{Grosshans2003b}. By equivalence we mean that the mutual information shared between Alice and Bob is the same in the two schemes and that the signal mode through the quantum channel looks the same to an outside observer such as Eve in both schemes.

The symplectic eigenvalues of (\ref{eq:matrix})~\cite{Serafini2004}, denoted $\nu_{AB,+}$ and $\nu_{AB,-}$, can now be used to find the entropy $S(AB)$ via the relation $S(AB) = g(\nu_{AB,+}) + g(\nu_{AB,-})$ where $g$ is the bosonic information function, $g(x) = \frac{x+1}{2} \log_2 \left(\frac{x+1}{2} \right) - \frac{x-1}{2} \log_2 \left( \frac{x-1}{2} \right)$~\cite{Holevo1999}. Likewise we find the conditional entropy $S(A|B)$ from the symplectic eigenvalue, $\nu_{A|B}$, of the conditional covariance matrix $\gamma_{A|B} = \gamma_A - \gamma_{B,11}^{-1} \sigma_{AB} \Pi \sigma_{AB}$, where $\Pi = \text{diag}[1,0]$ assuming an $X$-quadrature measurement at Bob, and $\gamma_{B,11}$ is the first diagonal element of $\gamma_{B}$. It follows then that $S(A|B) = g(\nu_{A|B})$. 

Finally, we arrive at the condition, $g(\nu_{AB,+}) + g(\nu_{AB,-}) = g(\nu_{A|B})$, for the complete elimination of Holevo information between Eve and Bob. For more details on this derivation, see the Supplementary Information~\cite{SI}. For a purely lossy channel without any excess noise ($\epsilon = 0$) this translates into the simple relation: $V_\text{sqz} + V_\text{sig} = 1$. This implies that $\chi_{EB}$ can become zero while $R \neq 0$. It is clear that this relation cannot be realized with coherent states as in this case $V_\text{sqz} = 1$ thus rendering the alphabet of zero size; $V_\text{sig}=0$. Squeezed states for which $V_\text{sqz} < 1$ are thus required to eliminate the Holevo quantity. To fulfill the condition, the size of the Gaussian alphabet has to be $V_\text{sig}=1-V_\text{sqz}$, and for very large squeezing degrees ($V_\text{sqz} \to 0$) the secure information rate in (\ref{eq:rate}) approaches $R = \beta I_{AB} = -\beta\frac{1}{2} \log_2(1-\eta)$. This shows that a secret key can in principle be generated for any channel loss and for any post-processing efficiency. It is also interesting to note that the elimination of the Holevo information is completely independent on the noise in the anti-squeezed quadrature, that is, it is independent on the impurity of the squeezed states \cite{Usenko2011}. We further remark that for ideal reconciliation efficiency, $\beta = 1$, the rate reaches half of the fundamental repeaterless bound for which $R = -\log_2(1-\eta)$~\cite{Pirandola2017}.

\begin{figure}
\includegraphics[width=0.70\textwidth]{
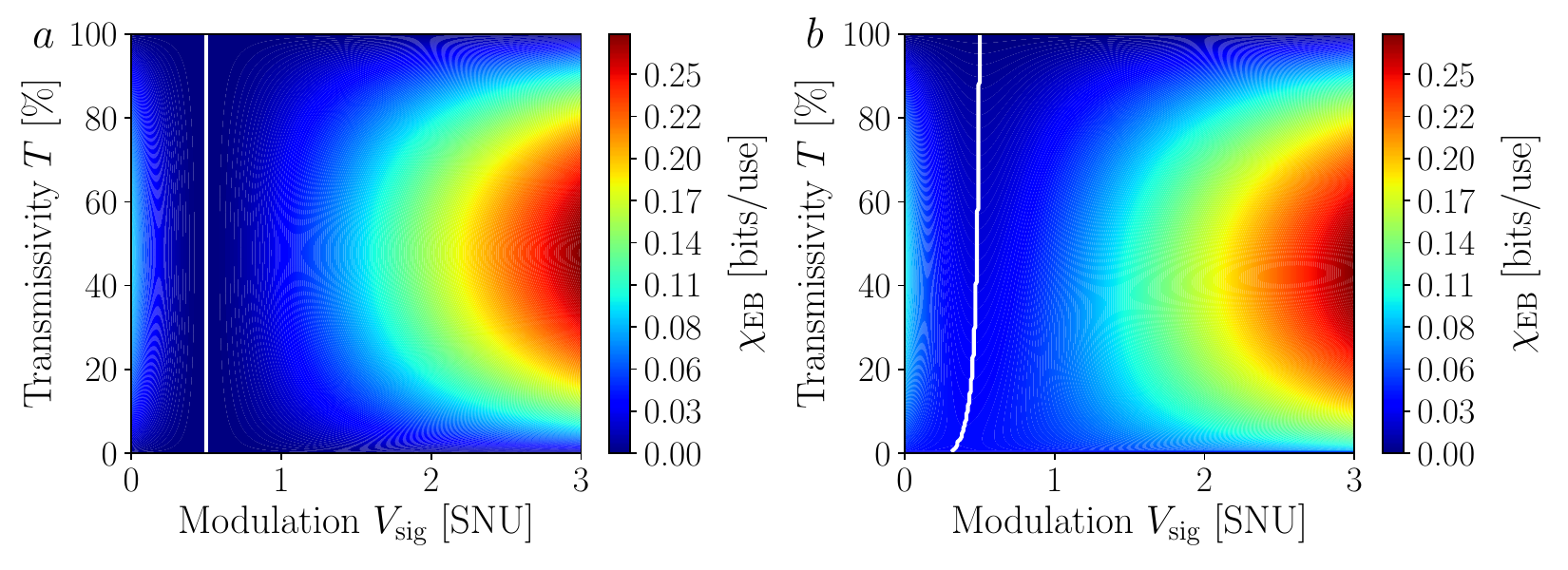}
\caption{Numerical calculation of leakage elimination. (a) Contour plot of the Holevo information bound in terms of signal modulation and transmittance in the quantum channel, with no excess noise and 0.5 SNU squeezing, corresponding to -3 dB. The white line indicates the minimum information leakage. (b) Contour plot of the Holevo information bound in terms of signal modulation and transmittance loss in the quantum channel, with excess noise $\epsilon = 0.01$ SNU and 0.5 SNU squeezing, corresponding to -3 dB. The white line indicates the minimum, but non-zero, information leakage.}
\label{fig:HolevoTheoryPlots}
\end{figure}

Evaluation of the Holevo quantity for the general case is found numerically and is shown in Fig. \ref{fig:HolevoTheoryPlots} for a purely lossy channel (Fig. \ref{fig:HolevoTheoryPlots}(a)) and for a channel with an untrusted excess noise of $\epsilon = 0.01$ shot-noise units (SNU) (Fig. \ref{fig:HolevoTheoryPlots}(b)). The minima of the 
Holevo information are marked by the white curves which for the purely lossy channel is exactly zero ($\chi_{EB}=0$) regardless of the transmittance for $V_\text{sig} = V_\text{sqz}=0.5$ SNU.    

While proper state modulation can eliminate the Holevo information between Eve and Bob, it does not eliminate the quantum mutual information between them, defined as $S(E) + S(B) - S(EB)$. This means that the subsystems $E$ and $B$ remain correlated in the quantum sense despite the fact that the information leakage is terminated. Such quantum mutual information vanishes completely only when no squeezing and no modulation is realized by the sender, which is shown in detail in the Supplementary Information~\cite{SI}. We also note that the correlations remain non-zero in the conjugate quadrature, but this is irrelevant since information is only encoded in the amplitude quadrature. Though single quadrature encoding reduces the alphabet, it does not compromise security, once basis switching and channel estimation are performed. In an actual implementation the conjugate quadrature would have to be measured to check the magnitude of the excess noise~\cite{Usenko2015, Gehring2016}.

The obtained result is based on the security analysis of Gaussian CVQKD protocols against collective attacks, which has been shown to be valid against the most general coherent attacks in the asymptotic limit \cite{Renner2009}. The estimation of the lower bound on the key rate is thus performed in the asymptotic regime. In the finite-size regime the lower bound on the key rate is further decreased by the security parameter $\Delta$ \cite{Leverrier2010}, which depends on the failure probability of the privacy amplification and speed of convergence of smooth min-entropy to von Neumann entropy \cite{Furrer2011}. For finite data, the minimization of information leakage becomes even more important, allowing trusted parties to partly compensate the reduction of the key rate due to finite-size effects, using proper state engineering, which does not affect the implementation-dependent $\Delta$ parameter directly.

\subsection*{Generation of states with no information leakage} 

\begin{figure}
\includegraphics[width=0.4\textwidth]{
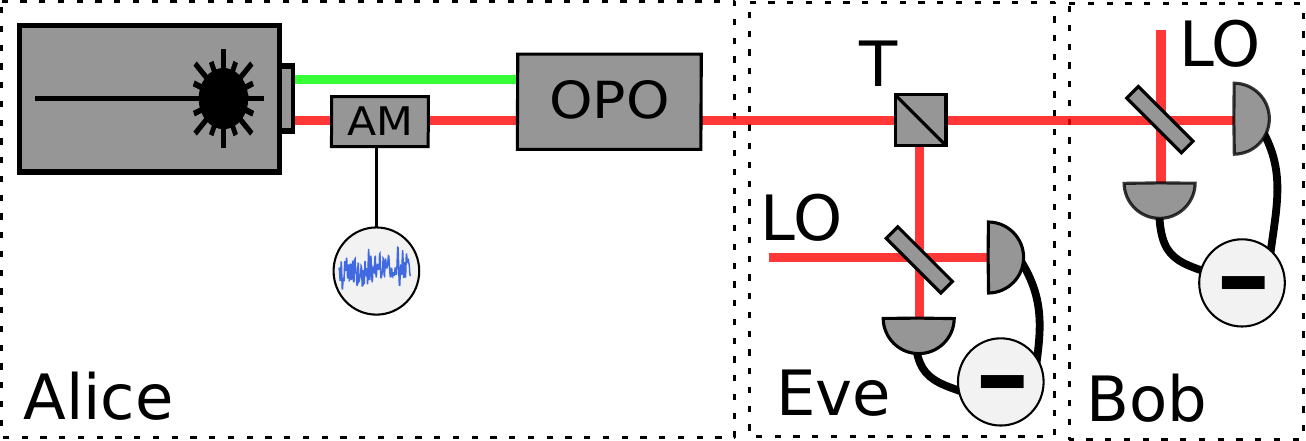}
\caption{Scheme of the experimental implementation of the quantum communication protocol with zero information leakage. An ensemble of coherent states is prepared by adding white noise to an amplitude modulator. The ensemble is squeezed by injecting it into an OPO, which is pumped by light at $\lambda/2$. The resulting ensemble of squeezed coherent states is sent to the quantum channel, with transmissivity $T$. Bob measures the channel output using homodyne detection, using an LO from the same laser. Eve also uses homodyne detection with an LO from the same laser. AM: Amplitude Modulator, OPO: Optical Parametric Oscillator, LO: Local oscillator, T: Transmissivity of the quantum channel.}
\label{fig:SetupSketch}
\end{figure}

We now implement a proof of principle experiment 
demonstrating the complete elimination of the information to an eavesdropper in a lossy channel. 
A schematic of the setup is depicted in Fig.~\ref{fig:SetupSketch}. The state is produced experimentally by squeezing an asymmetric thermal state: A bright laser beam at 1064 nm 
is modulated using an electro-optical modulator that is driven by a function generator. It produces white 
noise within the detection bandwidth, and forms sidebands on the bright beam. These sidebands 
(at 4.9 MHz with a bandwidth of 90 kHz) carry the information and correspond to an asymmetric thermal state. 
The modulated light beam is subsequently injected into an optical parametric oscillator (OPO) which squeezes, 
in this case, the amplitude quadrature by 3 dB. For more details on the OPO we refer to \cite{Madsen2012}. The final output state is thus an asymetric squeezed state alphabet where the amplitude quadrature signal information 
is sent to a computer while the states are injected into the lossy transmittance channel. 
Channel loss is simulated by a beam splitter with controllable transmittances. Eve measures the amplitude and phase quadrature of the reflected part using a homodyne detector with an efficiency of 95~\%, while Bob uses a homodyne detector with 85~\% efficiency to measure the amplitude and phase quadratures of the transmitted part. The measured data are electronically down-converted to dc, low-pass filtered 
and digitized. We thus have access to the covariance matrices of Alice, Bob and Eve as well as the amplitude quadrature correlation coefficients between Alice and Bob and both quadrature correlation coefficients between Eve and Bob. By means of these entities, we are now in the position to estimate the Holevo information using two different approaches: Either conservatively assuming that Eve holds the entire purification of the state shared by Alice and Bob or, as a comparison, directly from Eve's measurements.

\subsubsection*{\textbf{Purification based estimation of Holevo information}} In the first approach to finding the Holevo information, Eve is powerful and thus holds the entire purification 
of the virtually entangled state shared between Alice and Bob, as is the case in a standard QKD analysis~\cite{Weedbrook2011}. In order to do this we need to perform the purification on Alice's site. This is done by transforming the measured parameters at Bob backward through the channel knowing its transmittance. This includes the amplitude quadrature correlations between Alice and Bob, $C_{AB,X}^{(0)}=C_{AB,X}/\sqrt{T}$, and the quadrature variances $V^{(0)}_{B,i}=(V_{B,i}+T-1)/T$ where $i={X,P}$. We are then in the position to construct the covariance matrix of the entanglement-equivalent scheme at Alice with the modes that we name 
$A$ and $A^{\prime}$. This state is then purified according to a 4-mode purification procedure based on the Bloch-Messiah reduction theorem \cite{Braunstein2005b}, also known as Euler decomposition~\cite{Weedbrook2011}, similarly to what was done in \cite{Madsen2012}. 
The result of this procedure is a pure state of 4-modes which we label $A A^{^\prime} C D$. Mode $A^{\prime}$ is then propagated through the channel to obtain the global state $A B C D$, which is then assumed to be purified by modes accessible only to Eve. Using this global state we finally calculate the Holevo information, and plot the result for different modulation strenghts and different transmittances as shown in Fig. \ref{fig:HolevoPlot} (blue dots).

\subsubsection*{\textbf{Direct estimation of Holevo information}} In the second approach we directly estimate the Holevo bound by performing homodyne detection on the mode of light reflected from the channel, which is accesible to Eve. We use the measured data at Eve and Bob as well as the correlation 
coefficients, to deduce their individual covariance matrices and the associated correlations. This allows us to simulate Eve's collective attack by finding the conditional von Neumann entropy $S(E|B)$ and Eve's von Neumann entropy $S(E)$. Finally, using relation (\ref{eq:Holevo}), we directly find the Holevo information and plot the results in Fig. \ref{fig:HolevoPlot} (red crosses) for different values for the modulation depth. 

\subsubsection*{\textbf{Theoretical prediction of Holevo information}}
In addition to the direct and purification based estimation of the Holevo information, we also plot the theoretically expected Holevo information in terms of the signal modulation in the prepare-and-measure scheme, by numerically evaluating Holevo information of the derived covariance matrix with the experimentally established channel parameters.

Complete elimination of the Holevo information for any of the realized transmittances is clearly visible at the previously established condition, namely for $V_\text{sig} = 0.5 \text{ SNU} = -3$ dB given a $V_\text{sqz} = 0.5 \text{ SNU} = -3$ dB squeezed state such that $V_\text{sig}+V_\text{sqz} = 1 \text{ SNU}$, regardless of the method used for the estimation. The direct estimation approach tends to underestimate the Holevo information, while the purification based approach closely follows the theoretical prediction of the entanglement-based scheme described previously. We provide further details on the three approaches in the Supplementary Information~\cite{SI}.

\begin{figure}
\includegraphics[width=0.48\textwidth]{
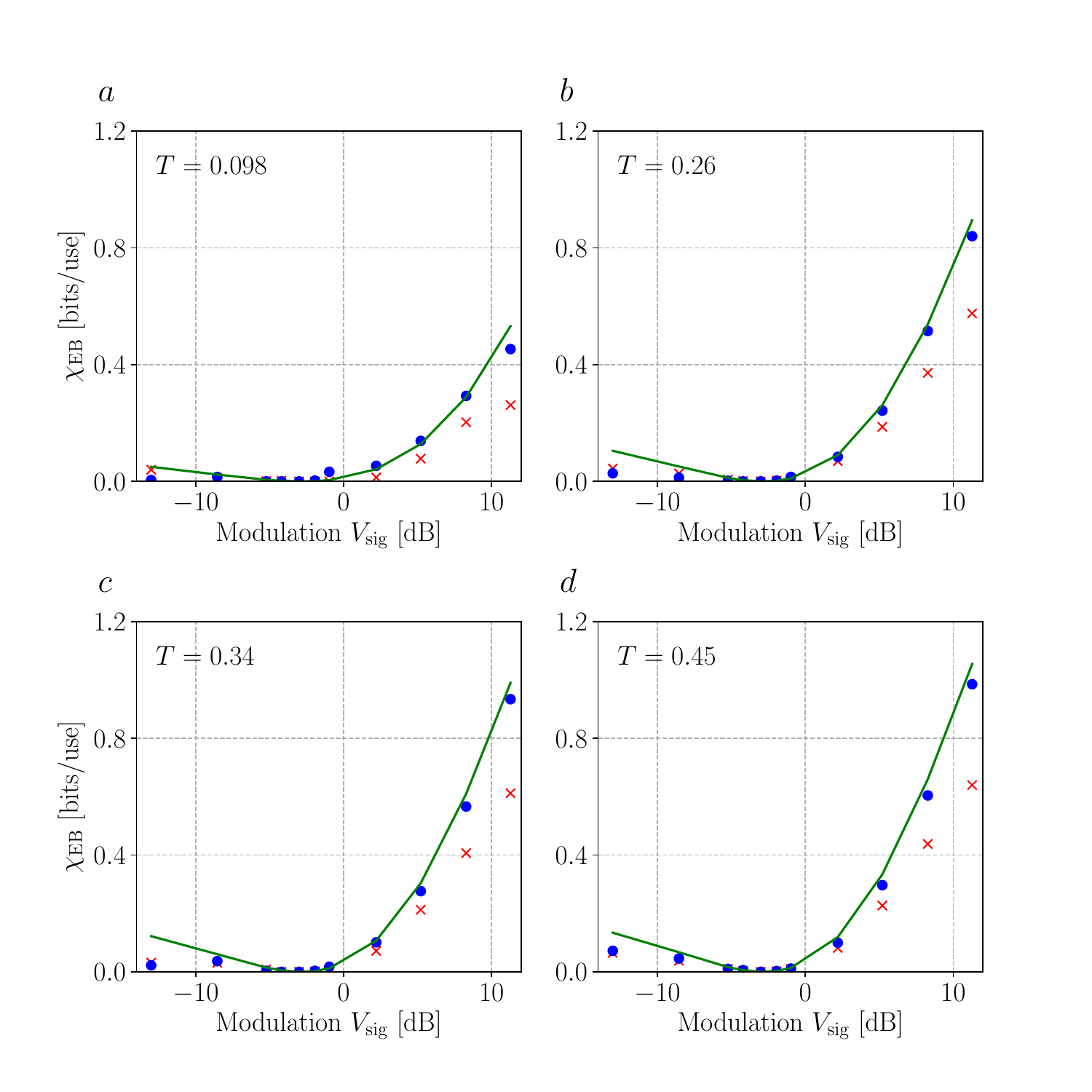}
\caption{Holevo information versus modulation depth for various transmittances. The modulation depth is normalized to the variance of shot noise. The Holevo information estimation was performed using three different approaches, namely direct estimation, general purification-based estimation, and a theoretical prediction from the channel parameters, shown with red crosses, blue dots and a green line respectively.}
\label{fig:HolevoPlot}
\end{figure}

It is evident from Fig. \ref{fig:HolevoPlot} that the direct estimation method underestimates the Holevo bound. This is caused by measurement imperfections that Eve ideally would not have. On the contrary, the purification-based approach corresponds to Eve perfectly obtaining maximum information associated with the noise level. This approach then closely follows the theoretical prediction obtained from the entangled-based scheme.

The measured data ensemble size of the order of $10^5$ was sufficiently large to provide good convergence of the Holevo bound and correspondence to the theory predictions. In a practical realization of QKD, however, the key is degraded by the finite-size effects and larger data ensemble sizes would be required to make this effect negligible. We estimate the value of the $\Delta$ parameter~\cite{Leverrier2010} in the Supplementary Information~\cite{SI}.

It is worth mentioning that the aim of our experiment is not to produce a secret key, but to demonstrate the complete elimination of the Holevo information in a purely lossy channel. To produce a secret key, it is important to implement random detection of conjugate quadratures at Bob's station and to modulate the anti-squeezed quadrature at Alice's station for increased key rate or use a slightly modified analysis assuming single quadrature modulation~\cite{Usenko2015, Gehring2016}.

\section*{Discussion}

In our study we first considered purely lossy channels, in which complete elimination of information leakage can be achieved. Noise may appear first of all as the result of imperfect detection, but in this case it can be calibrated and assumed trusted, i.e. being out of control by Eve. Since such noise does not contribute to Eve's information on Bob's measurement results~\cite{Usenko2016}, the complete elimination of information leakage can also be achieved upon detection noise using the same modulation setting. In the case when untrusted noise is present in the channel, however, the information leakage to Eve cannot be completely eliminated, but it can be effectively minimized using the same condition on state preparation as for a lossy channel~\cite{Usenko2011}. 

We have shown theoretically that a properly modulated squeezed state can be used to completely and deterministically decouple an eavesdropper from a purely lossy quantum channel without the use of entanglement distillation. The scheme has been confirmed experimentally using squeezed states of light and homodyne detection. The decoupling was shown to be completely independent on the amount of losses in the channel and the purity of the squeezed states used in the alphabet. This result is of fundamental interest in the context of quantum security, and we believe that the proposed protocol could offer an advantage, particularly in conjunction with a simple Gaussian error correcting scheme such as~\cite{Lassen2013} for the removal of non-Markovian excess noise, with channel multiplexing~\cite{Filip2005} or increased repetition rate.

A direction of further study can be the application of our proposed scheme in CVQKD with low efficiency error correcting codes, where an overall speedup in secret key generation may result from a partial reduction of Eve's information, even though the size of the alphabet is reduced. This can be useful in schemes where the error correction step limits the key generation rate.

\section*{Methods}

We refer to the Supplementary Information~\cite{SI} for additional details on the derivation of the information elimination condition and experimental methods.

\section*{Competing Interests}

The Authors declare no competing financial or non-financial interests.

\section*{Contributions}

R.F and V.C.U. developed the theory. L.S.M. and U.L.A. conceived the experiment. C.S.J and L.S.M. performed the experiment. C.S.J, V.C.U and L.S.M. analyzed the data. All authors contributed to the manuscript.

\section*{Funding}

We acknowledge support from the Center for Quantum Innovation (Qubiz), supported by the Innovation Foundation Denmark. We also acknowledge support from the Danish National Research Foundation (bigQ). V.C.U. acknowledges the project LTC17086 of the INTER-EXCELLENCE program of the Czech Ministry of Education, project 7AMB17DE034 of the Czech Ministry of Education and COST Action CA15220 "QTSpace". R.F. acknowledges grant No. 18-21285S of the Czech Science Foundation.

\section*{Data availability}

Raw data and scripts for the computation of the Holevo quantity are available from the authors upon reasonable request.

\section*{Supplementary Information}

Supplementary information is available at npj Quantum Information’s website.

\section*{References}


\newpage

\end{document}